\documentclass[10pt, oneside]{article} 
\usepackage{graphicx}			
\usepackage{amsmath}    
\usepackage{verbatim}    
\usepackage{subfigure}  
\usepackage{cite}

\usepackage{amssymb}
\usepackage{ulem}
\usepackage{authblk}

\usepackage{amssymb}
\usepackage{graphicx}
\usepackage{wasysym}
\usepackage{mathtools}

\renewcommand{\b}[1]{{\bf #1}}
\newcommand{\mrm}[1]{\mathrm{#1}}

\def\ba#1\ea{\begin{align}#1\end{align}}
\def\bg#1\eg{\begin{gather}#1\end{gather}}
\def\bpm{\begin{pmatrix}}
\def\epm{\end{pmatrix}}

\begin{document}

\title{Singular Flat Bands}

\author[1,2]{Jun-Won Rhim\thanks{Corresponding author: phyruth@gmail.com}}
\author[,1,2,3]{Bohm-Jung Yang\thanks{Corresponding author: bjyang@snu.ac.kr}}

\affil[1]{Center for Correlated Electron Systems, Institute for Basic Science (IBS), Seoul 08826, Korea}
\affil[2]{Department of Physics and Astronomy, Seoul National University, Seoul 08826, Korea}
\affil[3]{Center for Theoretical Physics (CTP), Seoul National University, Seoul 08826, Korea}

\begin{abstract}
We review recent progresses in the study of flat band systems, especially focusing on the fundamental physics related to the singularity of the flat band's Bloch wave functions.
We first explain that the flat bands can be classified into two classes: singular and non-singular flat bands, based on the presence or absence of the singularity in the flat band's Bloch wave functions.
The singularity is generated by the band crossing of the flat band with another dispersive band. 
In the singular flat band, one can find special kind of eigenmodes, called the non-contractible loop states and the robust boundary modes, which exhibit nontrivial real space topology.
Then, we review the experimental realization of these topological eigenmodes of the flat band in the photonic lattices.
While the singularity of the flat band is topologically trivial, we show that the maximum quantum distance around the singularity is a bulk invariant representing the strength of the singularity which protects the robust boundary modes. 
Finally, we discuss how the maximum quantum distance or the strength of the singularity manifests itself in the anomalous Landau level spreading of the singular flat band when it has a quadratic band-crossing with another band.
\end{abstract}

\maketitle

\section{Introduction}
A flat band indicates a type of band structures with constant energy independent of the crystal momentum.
The key feature of the flat band is that the charge carriers in it have a zero group velocity and an infinite effective mass.
Due to the suppressed kinetic energy, flat band systems have been considered intriguing in various research areas~\cite{sutherland1986localization,lieb1989two,arai1988strictly,aoki1996hofstadter,shima1993electronic,leykam2018artificial}. 
In condensed matter physics, flat bands are considered ideal to study many-body phenomena such as the ferromagnetism~\cite{mielke1993tasaki,tasaki1998nagaoka,mielke1999stability,hase2018possibility,you2019flat}, superconductivity~\cite{volovik1994fermi,yudin2014fermi,volovik2018graphite,aoki2020theoretical,kononov2020flat}, and Wigner crystal formation~\cite{wu2007flat,chen2018ferromagnetism,jaworowski2018wigner} because the kinetic energy of the carriers in the flat band is quenched and dominated by the electron-electron interaction.
In bosonic systems such as the photonic crystals, the realization of the slow light via a flat band has gotten a great attention because it can be applied to enhance the light-matter interaction~\cite{gersen2005real,baba2007slow,krauss2007slow,settle2007flatband,baba2008slow,corcoran2009green,schulz2010dispersion}.

Usually, we need a fine tuning of the system parameters to have a flat band, and it has hampered the experimental realization of such flat band.
However, there have been considerable experimental efforts to synthesize real materials or artificial systems with a nearly flat band.
In the condensed matter physics community, nearly flat bands are discovered in various kagome-type materials such as CoSn~\cite{kang2020topological,liu2020orbital}, Co$_3$Sn$_2$S$_2$~\cite{yin2019negative}, FeSn~\cite{kang2020dirac}, Fe$_3$Sn$_2$~\cite{ye2018massive,lin2018flatbands}, and YCr$_6$Ge$_6$~\cite{yang2019evidence,wang2020experimental}, as well as pyrochlore oxides~\cite{hase2018possibility,hase2019flat}.
Also, by using the novel band structure engineering technique, so-called twistronics, many nearly flat band systems have been realized in various Moire superlattice systems composed of misaligned two-dimensional layers, such as the twisted bilayer graphene~\cite{cao2018unconventional,cao2018correlated,yoo2019atomic,xie2019spectroscopic,kerelsky2019maximized,utama2020visualization}, twisted bilayer transition metal dichalcogenides~\cite{wang2020correlated,zhang2020flat}, and twisted multi-layer silicene~\cite{li2018realization}.
In particular, the twisted bilayer graphene at the magic angle~\cite{kim2017tunable} has attracted a great attention due to its unconventional superconductivity~\cite{liu2018chiral,cao2018unconventional,wu2018theory,po2018origin,yankowitz2019tuning,lian2019twisted,roy2019unconventional,gonzalez2019kohn,nunes2020flat,balents2020superconductivity,arora2020superconductivity,peri2020fragile} and Mott insulating phases~\cite{cao2018unconventional}.
Furthermore, in artificial systems, such as the photonic lattices~\cite{guzman2014experimental,vicencio2015observation,mukherjee2015observation_prl,xia2018unconventional,zong2016observation,xia2016demonstration,ma2020direct,mukherjee2015observation,weimann2016transport,travkin2017compact,cantillano2018observation,mukherjee2017observation,leykam2018perspective}, cold atom systems~\cite{jo2012ultracold,taie2015coherent}, engineered atomic lattices~\cite{drost2017topological}, and metamaterials~\cite{masumoto2012exciton,nakata2012observation,kajiwara2016observation,slot2017experimental,whittaker2018exciton}, one can have more controllability of the system parameters than the fermionic systems, and indeed almost flat bands were realized.

To understand the condensed matter phenomena properly, we usually start from the non-interacting physics of the given system in the band theory level, and then examine the effects of various interactions, such as the electron-electron interaction, based on the knowledge of the non-interacting electronic properties.
However, notwithstanding the rising interest in the interaction physics in flat bands discussed above, the nature of flat bands has not been well-understood even in the band theory level until recently.

In the view point of the topological band  theory, the flat band is simply considered trivial due to following reasons.
First, if the flat band is isolated from others and hopping processes are allowed within a finite range, the Chern number is always zero because the Bloch wave function is enforced to be analytic all over the Brillouin zone due to the flatness of the energy dispersion~\cite{chen2014impossibility,rhim2019classification}.
Second, even if the flat band is semimetallic by having a band-crossing point with another dispersive band, it cannot be a topological semimetal because the Berry phase obtained along a path enclosing the band-crossing point is usually unquantized and, even if it is quantized, its value is always $2n\pi$ (trivial) due to the band-flatness condition.~\cite{rhim2020quantum}.
Therefore, the flat band cannot exhibit the conventional bulk-boundary correspondence~\cite{hatsugai2009bulk,essin2011bulk,fukui2012bulk,graf2013bulk,bourne2015bulk,rhim2017bulk,rhim2018unified} such as the chiral edge modes from the nonzero Chern number in Chern bands~\cite{hatsugai1993chern,balents1996chiral,schulz2000simultaneous,kellendonk2002edge}, the Fermi arcs between two topological monopole charges with opposite signs in the Weyl semimetal~\cite{yang2011quantum}, and helical surface modes from Z$_2$ bulk topological invariant of topological insulators~\cite{kane2005quantum,kane2005z,bernevig2006quantum,fu2007topological,fu2007topological2}.

Although a flat band is topologically trivial, it was recently revealed that it can exhibit intriguing geometric properties.
To see what this means, let us consider the Hilbert-Schmidt quantum distance, which is defined as
\begin{align}
d^2(\psi_1,\psi_2) = 1 - \left| \langle \psi_1 | \psi_2\rangle \right|^2,\label{eq:hsqd}
\end{align}
where $\psi_1$ and $\psi_2$ are normalized quantum states~\cite{buvzek1996quantum,berry1989geometric,dodonov2000hilbert,rhim2020quantum}.
The quantum distance $d$, which takes a value between 0 and 1, measures how close the two states $\psi_1$ and $\psi_2$ are.
In the case of flat band systems, we consider the quantum distance between Bloch eigenstates in momentum space.
The quantum distance between two different Bloch states at momenta $\b k_1$ and $\b k_2$ is usually becomes zero when $|\b k_1 - \b k_2| \rightarrow 0$ because the two states take the identical form in this limit.
However, when a flat band has a band crossing with another dispersive band, the Bloch wave functions of the flat band around the band crossing point develop a nonzero quantum distance even if their momenta are close to each other.~\cite{rhim2019classification,rhim2020quantum}.
In this case, we call such a flat band the \textit{singular flat band}.
Interestingly, it is recently shown that the maximum value of the quantum distance among all the possible pairs of the Bloch eigenstates around the band-crossing point is a bulk invariant measuring the strength of the singularity at the band crossing point.~\cite{rhim2020quantum}.
Due to the singularity of the singular flat band, intriguing eigenmodes dubbed the \textit{non-contractible loop state} and the \textit{robust boundary mode} appear in systems with open boundaries and exhibit topological robustness in real space~\cite{rhim2019classification,ma2020direct}.
Moreover, the Landau levels of the singular flat band with a quadratic band-crossing show anomlous behavior, which is characterized by the maximum quantum distance~\cite{rhim2020quantum}.

These recent studies clearly demonstrate that the singular flat band is an intriguing new platform where the geometrical properties of Bloch states can be studied.
In general, the most crucial notions consisting of geometry are the curvature and metric (or distance).
In solids, Berry curvature and the quantum metric take the roles of the curvature and metric in the geometric description of the Bloch wave functions respectively, in momentum space.
While the curvature part has been studied exhaustively in past decades related to the semiclassical formulation of electron dynamics~\cite{xiao2010berry} and the topological classifications of materials~\cite{hasan2010colloquium}, physical implications of the metric or distance part have been much less investigated although there have been several studies on the physical roles of them in the current noise~\cite{neupert2013measuring}, the orbital magnetic susceptibility~\cite{raoux2015orbital}, the superfluid weight of a nearly flat Chern band~\cite{peotta2015superfluidity}, etc.
To achieve a unified understanding of the geometric nature of solids, further studies on the relations between the quantum metric or distance and physical quantities in various condensed matter phenomena have to be perfomed. 
The singular flat band model can be an ideal starting point of such research direction because this model's band-crossing point is characterized by the quantum distance instead of Berry curvature.

In this article, we pedagogically review the recent progress in the study of singular flat bands.
In particular, we focus on the novel geometrical properties arising from the fundamental relation between the singularity of the band crossing and the maximum quantum distance of a singular flat band. 
In Sec.~\ref{sec:flat_band_singularity}, we explain the relation between the existence of the singularity and the completeness of compact localized states, characteristic localized eigenstates of the flat band. 
Here we show that the nonzero maximum quantum distance guarantees the existence of non-local eigenstates, called the non-contractible loop state and the robust boundary mode, which induce intriguing topological features in real space.
In Sec.~\ref{sec:observation}, we briefly introduce the recent progress in the photonic lattice research, and then describe how the novel real-space topology was probed experimentally in the photonic lattices.
In Sec.~\ref{strength_of_singularity}, we show how to define the strength of the singularity of the singular flat band via the maximum quantum distance or the psedospin structure.
In Sec.~\ref{sec:anomalous_landau_level}, we explore how the maximum quantum distance is manifested in the exotic Landau level structure of the singular flat band with a quadratic band-crossing, which will pave the way for the measurement of the quantum distance of solids.
Here we also discuss how this phenomenon is related the the divergence of the orbital magnetic susceptibility.
Then, we show several candidate realistic materials, where those results can be observed.
Finally, in Sec.~\ref{sec:outlook}, we give conclusions and outlook of the research on the singular flat band systems.

\section{Flat band and singuarity}\label{sec:flat_band_singularity}

\begin{figure}
	\begin{center}
		\includegraphics[width=1\columnwidth]{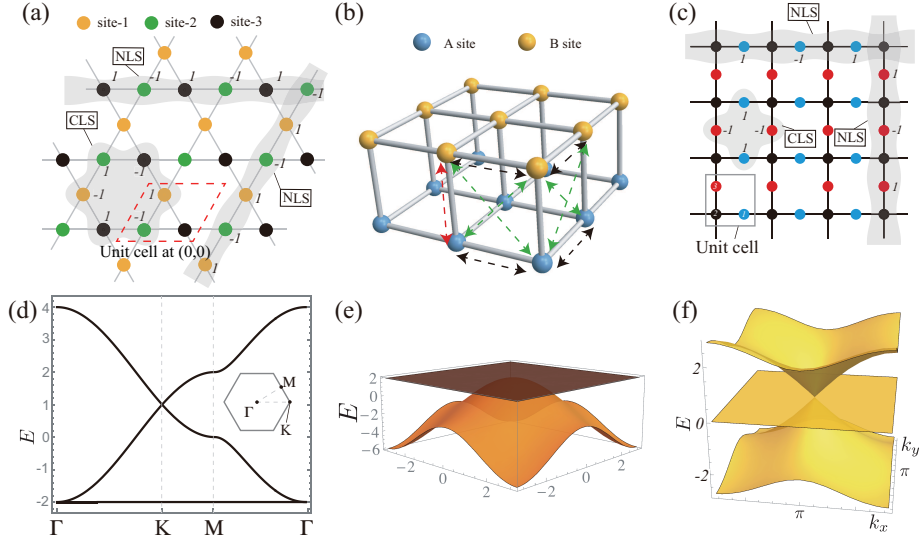}
	\end{center}
	\caption{ (a) A compact localized state (CLS) and two non-contractible loop states (NLSs) of the nearest neighbor tight binding model on the kagome lattice. The red dashed box represents the unit cell located at $\b R = (0,0)$. (b) The bilayer square lattice. Hopping parameters corresponding to the black and green dashed arrows are $1/2$, and those for the red arrows are $-2$. (c) A compact localized state and two non-contractible loop states of the Lieb lattice model with only the nearest neighbor hopping processes. Band structures for the models in (a), (b), and (c) are represented in the lower panels (d), (e), and (f) respectively. All the figures are adapted from \cite{rhim2019classification}.
	}
	\label{fig:cls_nls}
\end{figure}

\subsection{Compact localized state}\label{sec:cls}
The group velocity, the first derivative of an energy dispersion with respect to crystal momentum, represents the mobiility of the charge carriers in the band.
In the flat band, the group velocity is strictly zero for all momenta in the Brillouin zone.
This implies that charge carriers in the flat band are immobile or localized.
The localized nature of the flat band is explained by the existence of a special localized eigenstate called the compact localized state, whose amplitude is finite only inside a finite region in real-space, and exactly zero outside of it~\cite{sutherland1986localization,aoki1996hofstadter,rhim2019classification,bergman2008band,maimaiti2017compact,dubail2015tensor,read2017compactly}.
As a representative example, the compact localized state of the kagome lattice with the nearest neighbor hopping processes is shown in Fig.~\ref{fig:cls_nls}(a).
One can easily check that this compact localized state is an eigenstate of the kagome lattice model by applying the corresponding Hamiltonian operator to it.
Due to the destructive interference provided by the special lattice structure of the kagome lattice, which is also related to the geometrically frustrated lattice structure, the amplitudes of the compact localized state do not leak from its boundary, and the compact localized state recovers its original form after the hopping processes.
Note that the compact localized state is not unique, and one can have various types of the compact localized states by linear combinations of the smallest compact localized states centered at different positions.
Although the compact localized state looks similar to the Wannier function, they are quite different in the sense that (i) compact localized states usually do not form an orthonormal set while Wannier functions do, (ii) a compact localized state is an eigenmode while Wannier function is not usually.

\textit{A compact localized state is guaranteed to exist when the system has a flat band and the corresponding Hamiltonian is described by a tight binding model with a finite hopping range}~\cite{rhim2019classification}.
If the hopping processes are allowed within a finite range, each matrix element of the Bloch Hamiltonian $\mathcal{H}(\b k)$ of the tight binding model of such system is in the form of the finite sum of Bloch phases, which is given by
\begin{align}
\mathcal{H}(\b k)|_{lm} = \sum_{n_1 = p^{lm}_{1} }^{ q^{lm}_{1} } \cdots  \sum_{n_d = p^{lm}_{d} }^{ q^{lm}_{d} }  h_{lm}(n_1,\cdots,n_d) e^{in_1 k_1}\cdots e^{in_d k_d},
\end{align}
where $d$ is the dimension of the system described by $d$ primitive vectors $\b a_i$, $k_i = \b k \cdot \b a_i$, $n_i$ is an integer ranging from $p^{lm}_i$ to $q^{lm}_i$, and $h_{lm}(n_1,\cdots,n_d)$ is a complex coefficient of the Bloch phase $e^{in_1 k_1}\cdots e^{in_d k_d}$.
In a simple manner, we denote the Bloch phas as $e^{i\b k\cdot\b R} = e^{in_1 k_1}\cdots e^{in_d k_d}$, where $\b R = (n_1,\cdots,n_d)$ is the lattice vector.
Then, the eigenvector $\b v_{\mrm fb}$ for the flat band with energy $\epsilon_{\mrm fb}$ is obtained by solving coupled homogeneous equations given by $[\mathcal{H}(\b k) - \epsilon_{\mrm {fb}} ]\b v_{\mrm{fb}}(\b k) = 0$.
Since all the matrix elements of $\mathcal{H}(\b k) - \epsilon_{\mrm {fb}}$ are also in the form of the finite sum of Bloch phases, one can always find a unnormalized solution for the $\b v_{\mrm {fb}}(\b k)$ in the form of the finite sum of Bloch phases.
By using the fact that any arbitrary linear combination of the Bloch wave functions in the flat band is also an eigenstate, one can construct an eigenstate expressed by
\begin{align}
|\chi_{\b R}\rangle = c_\chi \sum_{\b k \in \mrm BZ} \alpha_{\b k} e^{-i \b k\cdot \b R} |\psi_{\mrm {fb}}(\b k) \rangle,\label{eq:cls}
\end{align}
where $|\psi_{\mrm{fb}}(\b k) \rangle$ is the Bloch eigenstate of the flat band, $c_\chi$ is the normalization constant for $|\chi_{\b R}\rangle$.
Here, $\alpha_{\b k}$ is a mixing coefficient of the Bloch eigenstates with different $\b k$'s, in the form of a multiplication between $|\b v_{\mrm {fb}}(\b k)|$ and an arbitrary finite sum of Bloch phases.
Note that the Bloch eigenstate is in the form $|\psi_{\mrm {fb}}(\b k) \rangle = N^{-1/2} \sum_{\b R^\prime}\sum_{j=1}^Q e^{i\b k\cdot\b R^\prime} \tilde{v}_{\mrm {fb}}(\b k)|_j | \b R^\prime,j\rangle$, where $N$ is the total number of the unit cells in the system, $\tilde{v}_{\mrm {fb}}(\b k)|_j$ is the $j$-th component of the normalized eigenvector of the flat band given by $\tilde{\b v}_{\mrm {fb}} = \b v_{\mrm {fb}}/|\b v_{\mrm {fb}}(\b k)|$, $| \b R^\prime,j\rangle$ represents the $j$-th orbital in the unit cell located at $\b r=\b R^\prime$, and $Q$ is the size of the Bloch Hamiltonian.
As a result, the amplitude of the $j$-th orbital of $|\chi_{\b R}\rangle$ in the unit cell at $\b R^\prime$ is obtained as
\begin{align}
\langle \b R^\prime , j|\chi_{\b R}\rangle  = \frac{c_\chi}{\sqrt{N}} \sum_{\b k \in \mrm BZ} e^{-i\b k\cdot (\b R - \b R^\prime )} \frac{ \alpha_{\b k}  v_{\mrm {fb},j} }{ |\b v_{\mrm {fb}}(\b k)| }. \label{eq:cls_amp}
\end{align}
Since $\alpha_{\b k}  \b v_{\mrm {fb}} / |\b v_{\mrm {fb}}(\b k)|$ is in the form of the finite sum of Bloch phases, the value of $\langle \b R^\prime , j|\chi_{\b R}\rangle$ vanishes if $(\b R - \b R^\prime )$ is out of the range of the lattice vectors in the Bloch phases in $\alpha_{\b k}  \b v_{\mrm {fb}} / |\b v_{\mrm {fb}}(\b k)|$.
This implies that $|\chi_{\b R}\rangle$ is a compact localized state, and one can have various kinds of the compact localized state depending on the choice of $\alpha_{\b k}$.
This is also a systematic scheme to obtain the compact localized states from the Bloch eigenstate of the flat band.

Let us consider the kagome lattice model as an example.
Its Bloch Hamiltonian is given by
\begin{align}
\mathcal{H}_{\mrm {kagome}}(\b k) = \bpm 0 & 1+e^{-ik_3} & 1+e^{ik_2} \\  1+e^{ik_3} & 0 &  1+e^{-ik_1} \\  1+e^{-ik_2} &  1+e^{ik_1} & 0   \epm, \label{eq:kagome_ham}
\end{align}
where $\b a_1 = a(1,0)$, $\b a_2 = a(-1/2,\sqrt{3}/2)$, and $\b a_3 = -\b a_1 - \b a_2$.
If the unnormalized eigenvector for the flat band of this model is chosen to be $\b v_{\mrm {fb}}^{\mrm {kagome}} = (e^{ik_1}-1,1-e^{-ik_2},e^{-ik_2}-e^{ik_1})^{\mrm T}$, a compact localized state centered at $\b r = 0$ is obatined from (\ref{eq:cls_amp}) as
\ba
\bpm \langle \b R ,1|\chi_0^{\mrm {kagome}} \rangle  \\ \langle \b R ,2|\chi_0^{\mrm {kagome}} \rangle \\ \langle \b R ,3|\chi_0^{\mrm {kagome}} \rangle \epm = \frac{1}{\sqrt{6}}\bpm -\delta_{\b R,(-1,0)} +\delta_{\b R,(0,0)} \\ -\delta_{\b R,(0,0)} + \delta_{\b R,(0,1)} \\ -\delta_{\b R,(0,1)}  + \delta_{\b R,(-1,0)}  \epm.\label{eq:kagome_cls}
\ea
This is the compact localized state with the smallest size in the kagome lattice model as plotted in Fig.~\ref{fig:cls_nls}(a).
The $j$-th element of the column vector on the right-hand side of (\ref{eq:kagome_cls}) represents the amplitude of the compact localized state at the site-$j$, which is indicated by three colors in Fig.~\ref{fig:cls_nls}(a).
For example, the first element $-\delta_{\b R,(-1,0)} +\delta_{\b R,(0,0)} $ of this vector means that the amplitudes at the site-1 in the unit cells located at $\b R=(-1,0)$ and $\b R=(0,0)$ are the coefficients of $\delta_{\b R,(-1,0)}$ and $\delta_{\b R,(0,0)}$, namely -1 and 1 respectively.
In fact, these coeffecients can be inferred from the unnormalized form of the Bloch eigenstate $\b v_{\mrm {fb}}^{\mrm {kagome}}$ because there is an one-to-one correspondence between the coefficients of $e^{-i\b k\cdot\b R^\prime}$ in $\b v_{\mrm {fb}}^{\mrm {kagome}}$ and that of $\delta_{\b R,\b R^\prime}$ in (\ref{eq:kagome_cls}).
In general, one can easily derive the compact localized state from the unnormalized Bloch eigenstate in this way.

\subsection{Singular and non-singular flat bands}\label{sec:singular_nonsingular}
One fundamental question on the compact localized state is that can one find a complete set of $N$ compact localized states to span the whole flat band?~\cite{bergman2008band,rhim2019classification}
We need to answer this question to understand the localized nature of the flat band completely.
It seems quite obvious that $N$ compact localized states centered at $N$ different unit cells are linearly independent.
Here $N$ indicates the total number of unit cells in the system with a periodic boundary condition.
However, it was noted that, under the periodic boundary condition, those $N$ compact localized states fail to form a complete set sometimes~\cite{bergman2008band}.
Later, it was rigorously shown that \textit{one cannot find a complete set of compact localized states if the Bloch eigenstate of the flat band is discontinuous at a momentum in the Brillouin zone.}~\cite{rhim2019classification}
This singular point can only be generated by the band-crossing of the flat band with another dispersive band, and the Bloch eigenstate in an isolated flat band is always analytic over the whole Brillouin zone. 
This is consistent with the fact that the Chern number of an isolated flat band is always zero when the hopping range of the model is finite.~\cite{chen2014impossibility}.
A flat band with the singularity is called a \textit{singular flat band}, and otherwise, it is called a \textit{non-singular flat band}~\cite{rhim2019classification}.

The flat band of the kagome lattice model in Sec.~\ref{sec:cls} is an example of the singular flat band.
First, note that the flat band has a quadratic band-crossing at $\Gamma$ point ($\b k=0$).
We check whether the corresponding Bloch eigenstate is singular or non-singular at the band-crossing point as follows.
The normalized eigenvector of $\mathcal{H}_{\mrm {kagome}}(\b k)$ in (\ref{eq:kagome_ham}) is given by
\ba
\tilde{\b v}_{\mrm {fb}}^{\mrm {kagome}} = \frac{\b v_{\mrm {fb}}^{\mrm {kagome}}}{|\b v_{\mrm {fb}}^{\mrm {kagome}}|}  = \frac{1}{|\b v_{\mrm {fb}}^{\mrm {kagome}}|} \bpm e^{ik_1}-1 \\ 1-e^{-ik_2} \\ e^{-ik_2}-e^{ik_1} \epm,
\ea
where $|\b v_{\mrm {fb}}^{\mrm {kagome}}| = \{2 (3-\cos k_1 - \cos k_2 - \cos k_3) \}^{1/2}$.
Note that all the elements of the unnormalized eigenvector $\b v_{\mrm {fb}}^{\mrm {kagome}}$ vanish at $\Gamma$ point simultaneously.
Since the normalization factor $|\b v_{\mrm {fb}}^{\mrm {kagome}}|$ is also zero at $\Gamma$ point, every element of the normalized eigenvector $\tilde{\b v}_{\mrm {fb}}^{\mrm {kagome}}$ is in the form of zero over zero.
Therefore the value of $\lim_{\b k \rightarrow 0}\tilde{\b v}_{\mrm {fb}}^{\mrm {kagome}}$ depends on the choice of the path approaching $\Gamma$ point.
For instance, one can show that $\lim_{k_1 \rightarrow 0}\tilde{\b v}_{\mrm {fb}}^{\mrm {kagome}}(k_1,0) = (i,0,-i)/\sqrt{2}$ and $\lim_{k_2 \rightarrow 0}\tilde{\b v}_{\mrm {fb}}^{\mrm {kagome}}(0,k_2) = (0,i,-i)/\sqrt{2}$.
This implies that $\tilde{\b v}_{\mrm {fb}}^{\mrm {kagome}}$ is discontinuous at $\Gamma$ point, and therefore the flat band of the kagome lattice is singular.

Let us consider another example of the flat band with a band-crossing, the bilayer square lattice model described in Fig.~\ref{fig:cls_nls}(b).
In this model, only the nearest neighbor hopping processes with amplitude $1/2$ are allowed between the sites in the same layer, and the inter-layer hopping parameters corresponding to the red and green dashed lines are $-2$ and $1/2$ respectively.
The corresponding Hamiltonian is given by
\begin{align}
\mathcal{H}_{\mrm {BSL}} = \bpm \cos k_x + \cos k_y & \cos k_x + \cos k_y-2 \\ \cos k_x + \cos k_y-2 & \cos k_x + \cos k_y \epm.
\end{align}
Although the flat band has a band touching at $\Gamma$ point, it is a non-singular flat band because its eigenvector $\tilde{\b v}_{\mrm {fb}}^{\mrm {BSL}} = (1,1)^{\mrm T}/\sqrt{2}$ is just a constant and therefore analytic for all momenta.

As shown by these two examples, in general, we have to investigate the Bloch eigenstates of a flat band to determine whether the flat band is singular or not.
However, we note that a flat band of any one dimensional system is always non-singular irrespective of the presence or absence of band crossings~\cite{rhim2019classification}.

The most crucial result of the existence of the singularity in a flat band is that \textit{one cannot find a complete set of compact localized states spanning a singular flat band under the periodic boundary condition}~\cite{rhim2019classification}.
To show this, we denote $N$ number of different compact localized states (labeled by $j$) by
\ba
|\chi_{\b {R}_j}\rangle = c_\chi^{(j)} \sum_{\b k \in \mrm BZ} \alpha_{\b k}^{(j)} e^{-i \b k\cdot \b {R}_j} |\psi_{\mrm {fb}}(\b k) \rangle, \label{eq:cls_2}
\ea
which have different shapes and center positions depending on the choice of $\alpha_{\b k}^{(j)}$ and $\b {R}_j$ respectively.
Then, a determinant diagnosing their linear independence or dependence is given by
\begin{align}
D = \begin{vmatrix}
	\alpha_{\mathbf{k}_1}^{(1)} e^{-i\mathbf{k}_1\cdot\mathbf{R}_1} & \alpha_{\mathbf{k}_1}^{(2)} e^{-i\mathbf{k}_1\cdot\mathbf{R}_2}  & \cdots & \alpha_{\mathbf{k}_1}^{(N)} e^{-i\mathbf{k}_1\cdot\mathbf{R}_N} \\
	\alpha_{\mathbf{k}_2}^{(1)} e^{-i\mathbf{k}_2\cdot\mathbf{R}_1} & \alpha_{\mathbf{k}_2}^{(2)} e^{-i\mathbf{k}_2\cdot\mathbf{R}_2}  & \cdots & \alpha_{\mathbf{k}_1}^{(N)} e^{-i\mathbf{k}_2\cdot\mathbf{R}_N} \\
	\vdots  & \vdots & \ddots & \vdots \\
	\alpha_{\mathbf{k}_N}^{(1)} e^{-i\mathbf{k}_N\cdot\mathbf{R}_1} & \alpha_{\mathbf{k}_N}^{(2)} e^{-i\mathbf{k}_N\cdot\mathbf{R}_2}  & \cdots & \alpha_{\mathbf{k}_N}^{(N)} e^{-i\mathbf{k}_N\cdot\mathbf{R}_N}
\end{vmatrix},\label{eq:det_1}
\end{align}
because the Bloch wave functions $|\psi_{\mrm {fb}}(\b k) \rangle$'s with different momenta in (\ref{eq:cls_2}) form a complete set.
As noted in Sec.~\ref{sec:cls}, to ensure the compact localization of $|\chi_{\b {R}_j}\rangle$, $\alpha^{(j)}_{\b k}$ should be proportional to $|\b v_{\mrm{fb}}(\b k)|$, where $\b v_{\mrm{fb}}(\b k)$ is a unnormalized eigenvector of the flat band in the form of the finite sum of Bloch phases.
Therefore, one can note that
\ba
D \propto \prod_{l=1}^N |\b v_{\mrm{fb}}(\b k_l )|.
\ea
As a result, if $\b v_{\mrm{fb}}(\b k)$ is zero at a crystal momentum, namely the Bloch eigenstate $\tilde{\b {v}}_{\mrm{fb}}(\b k)$ is singular at that momentum, $D$ also becomes zero implying that $N$ compact localized states are always linearly dependent of each other.
As one can note from (\ref{eq:cls_2}), in this case, the Bloch wave function corresponding to the singular momentum is always absent in constructing the compact localized state because $\alpha^{(j)}_{\b k}$ is zero at that momentum.
Namely, $N$ compact localized states of a singular flat band are made of a smaller (less than $N$) number of Bloch wave functions, which implies that they cannot be linearly independent.
For example, in the kagome lattice, the $N$ compact localized states obtained by translating the minimal compact localized state in (\ref{eq:kagome_cls}) by all the lattice vectors are known to be linearly dependent of each other because a simple sum of all of them vanishes.
On the other hand, if the flat band is non-singular, namely there exists a nonzero $\b v_{\mrm{fb}}(\b k)$ for the flat band, one can obtain non-zero $D$ by choosing $\alpha^{(j)}_{\b k} =|\b v_{\mrm{fb}}(\b k)|$ and $\{\b R_j \}$ to be $N$ distinct lattice vectors.
Therefore, \textit{one can find a complete set of $N$ compact localized states for a non-singular flat band.}

We note that there is a subtle issue related to the finite size effect.
Let us denote the number of unit cells along $\b a_i$ direction by $N_i$, so that the total number of the unit cells in the system is given by $N = \prod_{i=1}^d N_i$.
Then, the crystal momentum discretized under periodic boundary condition along each reciprocal lattice vector is $k_i = (2\pi/N_i)m$, where $m$ is an integer.
Therefore, even though the continuum expression of the band dispersion shows a singular band-crossing at a certain crystal momentum $\b k_0$, one can find a complete set of $N$ compact localized states if $\b k_0$ cannot be represented by the form $k_i = (2\pi/N_i)m$ for the given system size $N_i$.
Let us consider the Lieb lattice model as an example.
Considering only the nearest neighbor hopping processes, the Hamiltonian is given by
\ba
\mathcal{H}_{\mrm{Lieb}} = \bpm 0 & 1+e^{ik_x} & 0 \\ 1+e^{-ik_x} & 0 & 1+e^{-ik_y} \\ 0 & 1+e^{ik_y}& 0 \epm.
\ea
The eigenvector of its flat band is obtained as
\ba
\tilde{\b v}_{\mrm{fb}}^{\mrm{Lieb}}= \frac{1}{\sqrt{4+2\cos k_x +2\cos k_y}}\bpm 1+e^{-ik_y} \\ 0 \\ -1-e^{-ik_x} \epm,
\ea
which is singular at $\b k_0 = (\pi,\pi)$.
Note that this $\b k_0 $ is allowed only when both $N_x$ and $N_y$ are even numbers.
If we choose a compact localized state for this model as shown in Fig.~\ref{fig:cls_nls}(c), any linear combination of the $N=N_x N_y$ translated copies of this compact localized state is nonzero for odd values of $N_x$ and $N_y$.
This implies that these compact localized states form a complete set.
On the other hand, if $N_x$ and $N_y$ are even integers, the same staggered sum of them vanishes.
In this case, these compact localized states are linearly dependent on each other.
Note that if the singular band-crossing is at $\b k=0$, we are free from this finite size issue in general.

\subsection{Non-contractible loop and planar states}
Since we cannot construct a complete set of basis wave functions for the singular flat band from the compact localized states as discussed in Sec.~\ref{sec:singular_nonsingular}, some non-compact eigenstates should be included in addition to the compact localized states to span the flat band completely.
In the kagome lattice, as an example, such non-compact modes were found to be the so-called \textit{non-contractible loop states}, which are shown in Fig.~\ref{fig:cls_nls}(a)~\cite{bergman2008band}.
They are extended at least along one spatial direction while compactly localized along other directions.
One can check that there are two such extended eigenmodes in the kagome lattice under the periodic boundary condition, which are independent of the compact localized states.
Any other non-contractible loop states at different positions and shapes can be constructed by a linear combination of the given non-contractible loop states and the compact localized states.
In the kagome lattice case, it was noted that $N-1$ number of compact localized states can be linearly independent of each other~\cite{bergman2008band}.
Therefore, the existence of these two additional non-contractible loops states is consistent with the fact that there are $N+1$ number of degenerate Bloch eigenstates at the energy of the flat band, which indicates that the flat band should be touching with another dispersive band at a momentum.

The non-contractible loop states exhibit intriguing topological aspects in real space as follows.
For a two-dimensional system under periodic boundary condition, the two independent non-contractible loop states form two closed loops encircling the torus geometry along the toroidal and poloidal directions, respectively.
Also, they are robust because a non-contractible loop state cannot be cut in the middle by any linear combination between the non-contractible loop state and compact localized states, which merely deforms the shape of the non-contractible loop state.

In general, it was shown that such non-contractible loop states are guaranteed to exist in a two dimensional singular flat band~\cite{rhim2019classification}.
If the band crossing point is at $\b k=\b k_0$, the Bloch wave function corresponding to this momentum does not participate in constructing a compact localized state by the linear combination of Bloch wave functions as discussed previously via the equation (\ref{eq:cls_2}).
As a result, any kind of eigenmodes including the Bloch wave function at $\b k=\b k_0$ is linearly independent of the compact localized states.
While the Bloch wave function at $\b k=\b k_0$ is extended along both directions $\b a_1$ and $\b a_2$, one can make it compactly localized along one of those two directions by a linear combination of it with other Bloch wave functions as follows.
Let $\mathcal{H}(k_1,k_2)$ be the Hamiltonian under consideration.
If we fix $k_1$ to $k_{0,1}$, where $\b k_0 = (k_{0,1},k_{0,2})$, we obtain an effective one dimensional flat band Hamiltonian $\mathcal{H}(k_{0,1},k_2)$ with momentum $k_2$.
Since any one dimensional flat bad is a non-singular flat band as noted previously, one can always find a compact localized state for this effective Hamiltonian localized along $\b a_2$ direction, which includes the Bloch wave function at $\b k=\b k_0$ too.
This is the non-contractible loop state of the full Hamiltonian $\mathcal{H}(k_1,k_2)$ because it is still extended along $\b a_1$ direction with momentum $k_{0,1}$.
In the similar way, one can find another non-contractible loop state extended along $\b a_2$ direction too.

In three dimensional case, one can have different kind of partially extended states with nontrivial real space topology, called the non-contractible planar states (NPSs)~\cite{rhim2019classification}.
If we denote a singular flat band Hamiltonian for this case by $\mathcal{H}(k_1,k_2,k_3)$, one should fix two components of $\b k = (k_1,k_2,k_3)$ to the corresponding components of the singular momentum $\b k_0 = (k_{0,1},k_{0,2},k_{0,3})$ to obtain an effective one dimensional flat band Hamiltonian in this case.
Let us assume that those two fixed momenta are $k_1=k_{0,1}$ and $k_2 = k_{0,2}$, for convenience.
Then, we can construct a planar eigenstate, namely a NPS, which is extended along two directions $\b a_1$ and $\b a_2$, and compactly localized along $\b a_3$.
By fixing other pairs of momentum components, one can obtain other NPSs along different directions.

\begin{figure}
	\begin{center}
		\includegraphics[width=1\columnwidth]{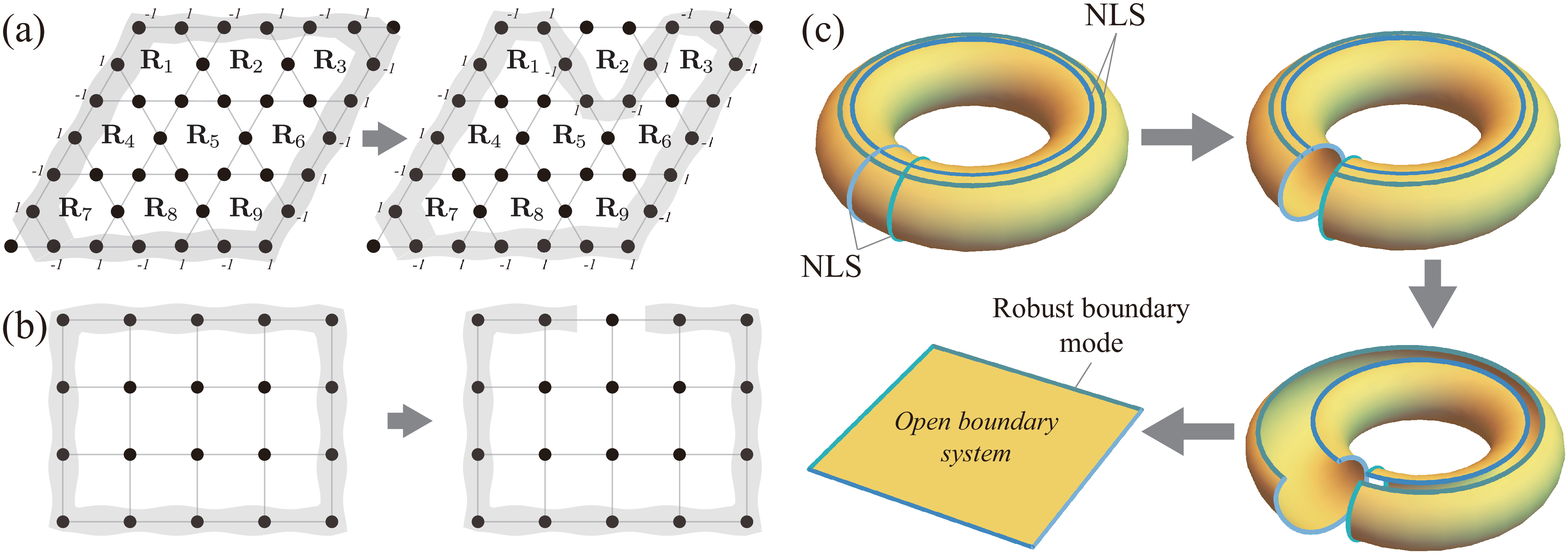}
	\end{center}
	\caption{ (a) The robust boundary mode of the kagome lattice model. This can be obtained by adding all the compact localized states centered at all the unit cells (from $\b R_1$ to $\b R_9$). If one subtract a compact localized state centered at $\b R_2$ to try to cut the robust boundary mode, it cannot be cut but deformed as illustrated in the right-hand side. (b) A boundary eigenmode on the bilayer square lattice, which is obtained by adding compact localized states centered at boundary sites. This boundary mode can be cut by subtracting a compact localized state on the boundary as shown in the right-hand figure. In (c), it is illustrated schematically how the non-contractible loop states are connected to the robust boundary mode by deforming the periodic boundary condition to the open boundary condition. All the figures are from \cite{rhim2019classification}
	}
	\label{fig:rbm}
\end{figure}

\subsection{Robust boundary modes}
Under the open boundary condition, the singularity of a singular flat band is manifested by the existence of a boundary eigenmode, called the robust boundary mode~\cite{rhim2019classification}.
The robust boundary mode of the kagome lattice is shown in Fig.~\ref{fig:rbm}(a) as an example.
In fact, the robust boundary mode is closely related to the non-contractible loop state because the non-contractible loop states can transform to the robust boundary mode as we modify the periodic boundary condition to the open boundary condition as follows.
We first prepare four non-contractible loop states on a torus geometry of the system respecting the periodic boundary condition, two along the toroidal direction and the other two along the poloidal direction.
Then, we transform the torus into a finite system with the open boundary condition by cutting the surface of the torus between non-contractible loop states as illustrated in Fig.~\ref{fig:rbm}(c)
In this process, the four non-contractible loop states lead to the robust boundary mode.
Since the appearance of the non-contractible loop states is from the singularity of the singular flat band, the existence of the robust boundary mode is also due to the same singularity.

This is a new kind of bulk-boundary correspondence between the singularity in the bulk system and the boundary mode of the finite system.
The conventional bulk-boundary correspondence in topological systems is the correspondence between a bulk topological invariant and the in-gap boundary modes of the finite system.
Examples of the topological invariant are the Chern number of the Chern bands or Landau levels~\cite{hatsugai1993chern,balents1996chiral,schulz2000simultaneous,kellendonk2002edge}, Z$_2$ index in topological insulators~\cite{kane2005quantum,kane2005z,bernevig2006quantum,fu2007topological,fu2007topological2}, the Zak phase of the one dimensional insulators with reflection symmetry~\cite{vanderbilt1993electric,rhim2017bulk,ryu2006entanglement,delplace2011zak}, and so on.
The robust boundary mode of the singular flat band is distinguished from the topological boundary modes in the following perspectives. 
First, the robust boundary mode is not in the bulk band gap but at the same energy of the flat band. 
Therefore, one cannot detect the robust boundary mode by electronic spectroscopy such as the angle-resolved photoemission spectroscopy (ARPES)~\cite{damascelli2003angle}. 
Instead, one should use other kinds of experimental tools, which can access the wave function directly. 
The photonic lattice set-up is appropriate for this because here wave functions can be prepared as well as detected quite freely.
Second, the robust boundary mode is protected by the singularity and the flatness of the band while the topological edge states are protected by the topological charge or certain crystal symmetries.
Third, robust boundary mode is linearly dependent of the compact localized states of the bulk, while the topological edge modes are not.
Note that compact localized states are linearly independent of each other under the open boundary condition, and form a complete set spanning the flat band.
It is worth noting that the compact localized states of a singluar flat bands are linearly dependent so that their linear combination becomes zero under periodic boundary condition, whereas the same linear combination generates a robust boundary mode under open boundary condition.
Since the robust boundary mode is obtained from a macroscopic number, corresponding to the system size, of the compact localized states, it cannot be cut or destroyed by adding few number of compact localized states, and this is what we mean by the robustness.
Even in the non-singular flat band, one can construct a boundary eigenmode by collecting all the compact localized states spanning the boundaries of the finite system.
However, such boundary mode is not the robust boundary mode because it can be cut by just adding a compact localized state near the boundary as shown by the square lattice bilayer example in Fig.~\ref{fig:rbm}(b).
Let us note that the presence of robust boundary modes is confirmed in a recent experiment in photonic lattices as discussed in Sec.~\ref{sec:observation}.

\begin{figure}
	\begin{center}
		\includegraphics[width=0.5\columnwidth]{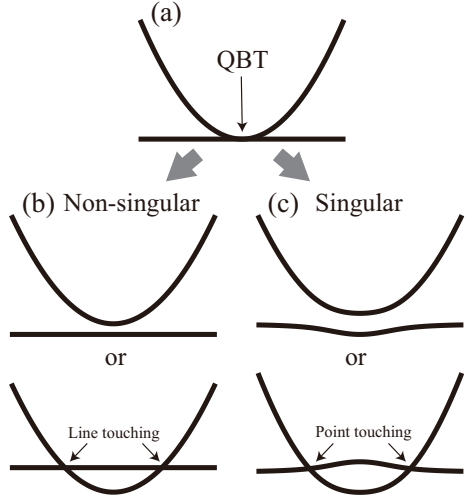}
	\end{center}
	\caption{ (a) A two-dimensional flat band with a quadratic band touching (QBT). (b) If the QBT is non-singular, one can gap out the QBT while maintaining the band flatness. When the flat band is shifted upward in the energy space, one obtains a line touching between the flat band and the parabolic band. (c) If the band-crossing point is singular, one should break the flatness of the flat band during the gap opening process of the QBT point. If the flat band is mixed with the quadratic band, one obtain point touching between two bands. This figure is adapted from \cite{rhim2019classification}
	}
	\label{fig:flatness_enforced}
\end{figure}

\subsection{Flatness enforced band-crossing}
Conventionally, a band-crossing is protected by crystalline symmetries or topological charge~\cite{chiu2016classification}.
Interestingly, in the case of the singular flat band, the protection mechanism of its singular band-crossing does not belong to any of them.
Instead, it was shown that it is protected by just the flatness of the flat band.
One can analyze the generic protection mechanism for the band-crossing of the flat band by using a low energy continuum Hamiltonian~\cite{rhim2019classification}.
For simplicity, we consider two dimensional systems with two bands touching each other.
If the flat band has a point touching with another band, the band-crossing type should be at least parabolic, because the energy dispersion of the band touching with the flat band becomes not differentiable in the case of the linear band-crossing.
The most general quadratic form of the two-band Hamiltonian is given by
\begin{align}
\mathcal{H}_{\b k} =& (t_1 k_x^2 + t_2 k_x k_y + t_3 k_y^2)\sigma_z + (t_4k_xk_y +t_5k_y^2)\sigma_y +t_6k_y^2\sigma_x \nonumber\\
& + (b_1 k_x^2 + b_2 k_x k_y + b_3 k_y^2)\sigma_0, \label{eq:quad_2by2}
\end{align}
where $\sigma_x$, $\sigma_y$, $\sigma_z$ are Pauli matrices, and $\sigma_0$ is the identity matrix.
Any quadratic Hamiltonian matrix can be unitarily transformed to the above form.
The condition for the existence of a flat band is given by
\begin{align}
\mathrm{det}\mathcal{H}_{\b k} = 0,
\end{align}
where we assume that the flat band is at the zero energy without loss of generality.
From this we obtain five constraints as follows.
\begin{align}
t_1^2 =& b_1^2, \label{eq:quad_1}\\
t_1t_2 =& b_1 b_2, \label{eq:quad_2}\\
t_3^2 + t_5^2 + t_6^2 =& b_3^2, \label{eq:quad_3}\\
t_2^2 +2 t_1t_3 + t_4^2 =& b_2^2 +2 b_1b_3, \label{eq:quad_4}\\
t_2t_3 + t_4t_5 =& b_2b_3. \label{eq:quad_5}
\end{align}
Under these constraints, the Hamiltonian $\mathcal{H}_{\b k}$ yields a flat band and a parabolic band touching each other at $\b k=0$.
One can note that the flat band is a singular flat band only when both $t_1$ and $t_4$ are nonzero.
The non-singular flat band and singular flat band show distinct features during the gap-opening process as follows.

First, the non-singular flat band can be decoupled from the parabolic band while maintaining the flatness of the energy dispersion.
In this case, it was shown that the Hamiltonian matrix $\mathcal{H}_{\b k}$ can be always transformed to $\tilde{\mathcal{H}}_{\b k} = (t_1^\prime k_x^2 + t_2^\prime k_x k_y + t_3^\prime k_y^2)(\sigma_z +\sigma_0) $.
Since $\tilde{\mathcal{H}}_{\b k}^\prime$ has no off-diagonal terms, a perturbation of the form $\mathcal{H}^\prime = \delta\lambda (\sigma_z-\sigma_0)$ can control the energy of the flat band without destroying its flatness.
As a result, we obtain a flatness-preserved gap opening or a line toughing between two bands when $\delta\lambda \neq 0$.
This is summarized in Fig.~\ref{fig:flatness_enforced} (b).

On the other hand, in the case of the singular flat band, the flat band cannot be detached from the parabolic band while maintaining its flatness when a mass term of the form $\mathcal{H}^\prime = m_x\sigma_x +  m_y\sigma_y + m_z\sigma_z + m_0\sigma_0$ is added to the system.
From the flatness conditions, $\mathrm{det}\mathcal{H}_{\b k} = 0$ and $\mathrm{det} \left( \mathcal{H}_{\b k} +\mathcal{H}^\prime \right)= 0$, we always obtain $t_4=0$ if one of the mass parameters is nonzero.
This contradicts with the fact that $t_4$ should be nonzero to have a singular flat band.
Therefore it is impossible to shift the energy of the singular flat band while preserving its flatness, which implies that the singular flat band's band-crossing is enforced by the band's flatness.
Depending on the choice of the mass parameters, one can either have the gap-opening at the band-crossing point or observe the generation of two Dirac points at different crystal momenta.
These results are illustrated in Fig.~\ref{fig:flatness_enforced}(c).
Moreover, one can find a mass term, which results in a Chern band separated from other bands.
As an example, in the kagome lattice, if we add a gap-opening perturbation $\mathcal{H}^\prime = \delta (\lambda_2+\lambda_7)$, where $\lambda_i$'s are the Gell-Mann matrices, to $\mathcal{H}_{\mathrm{kagome}}(\b k)$, we have a nearly flat band with a nonzero Chern number.
This can be a systematic route to obtain a nearly flat Chern band starting from a singular flat band when one uses a simple scheme to design a singular flat band model proposed in \cite{rhim2019classification}.

\section{Observation of the real-space topology in photonic lattices}\label{sec:observation}

\subsection{Flat bands in photonic lattices}
One can fabricate a two dimensional periodic array of optical waveguides, so-called the photonic lattice~\cite{efremidis2002discrete,fleischer2003observation,longhi2006observation,schwartz2007transport,peleg2007conical,lahini2008anderson,lederer2008discrete,chen2012optical,nixon2013observing,segev2013anderson,rechtsman2013strain,rechtsman2013topological,rechtsman2013photonic,plotnik2014observation,zeuner2015observation,el2019corner}, in a three dimensional medium by using the femtosecond laser writing technique~\cite{davis1996writing,pertsch2004discrete}.
A light beam propagating in this system satisfies a wave equation given by
\ba
-i\frac{\partial}{\partial z} \psi(x,y,z) = -\frac{1}{2k_0 n_0} \nabla_\perp^2\psi(x,y,z) -k_0 n(x,y) \psi(x,y,z),
\ea
within the paraxial approximaition~\cite{fleischer2003observation,lederer2008discrete,chen2012optical}, where $\psi(x,y,z)$ describes the electric field envelope of the light propagating along $z$-axis, $n_0$ and $n(x,y)$ are the refractive indices outside and inside the waveguide of the medium, $k_0$ is the wavenumber in vacuum, and $\nabla_\perp^2 = \partial_x^2 + \partial_y^2$.
This can be regarded as a 2D Schr\"{o}dinger equation by substituting $z$ with time, and $n(x,y)$ with the periodic potential.
Namely, the 2D cross section of the medium perpendicular to the $z$-axis corresponds to a 2D lattice model in solid state physics.
In the photonic lattice set-up, one can have more freedom to design lattice models, control band parameters of them, and prepare an initial state compared with the conventional condensed matter experiments.
As a result, many flat band systems such as the Lieb~\cite{guzman2014experimental,vicencio2015observation,mukherjee2015observation_prl,xia2018unconventional} and kagome lattice models~\cite{zong2016observation,xia2016demonstration,ma2020direct} have been studied by using the photonic lattice experimental set-ups~\cite{mukherjee2015observation,weimann2016transport,travkin2017compact,cantillano2018observation,mukherjee2017observation,leykam2018perspective}.

In the photonic lattice, a flat band is probed indirectly by showing that a compact localized state is an eigenmode because a compact localized state is guaranteed to exist in a flat band system and represents the localized nature of this dispersionless band.
If the waveguides embedded in a medium extend along $z$-direction from $z=0$ to $z=L_z$, an incident light beam, whose amplitudes and phases correspond to those of the compact localized state, enters into the waveguides at $z=0$.
As the light beam propagates along $z$-axis, it would also spread along the transverse direction in general due to the coupling between neighboring waveguides.
However, the light beam corresponding to the compact localized state remains same up to an overall phase factor during the propagation because the compact localized state is an exact eigenstate of the flat band system.
As a result, what we observe at the other end of the waveguides ($z=L_z$), is the same incident light beam only multiplied by a phase factor, and this is how one demonstrate the existence of a flat band in the photonic lattice experiments.

\begin{figure}
	\begin{center}
		\includegraphics[width=1\columnwidth]{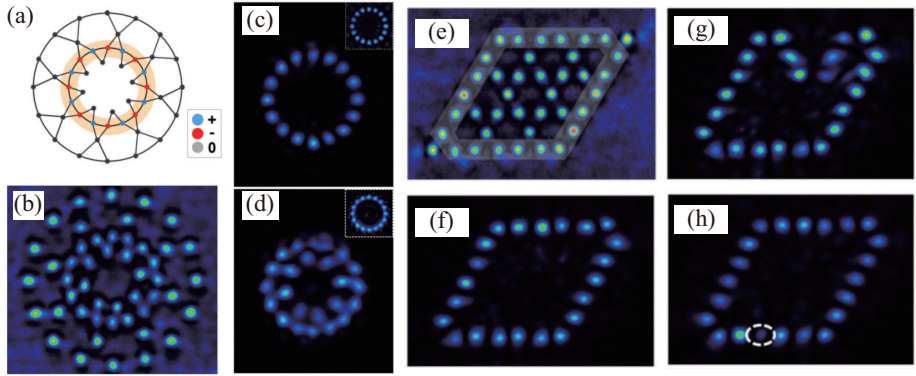}
	\end{center}
	\caption{ (a) The kagome lattice in the Corbino geometry. (b) Experimental realization of (a) in the photonic lattice. (c) Observation of the non-contractible loop state. With the out-of-phase configuration, the initial state propagates along $z$-direction without any transverse dissipation. (d) An initial state with the in-phase condition is not an eigenmode. (e) A finite kagome lattice with the open boundary condition. (f) Experimental realization of the robust boundary mode. The robustness of the robust boundary mode is tested by adding (g) a compact localized state, and (h) a defect. These figures are adapted from \cite{rhim2019classification}
	}
	\label{fig:observation}
\end{figure}

\subsection{Observing non-contractible loop states and robust boundary modes in photonic lattices}
As described in the previous section, non-contractible loop states and robust boundary modes should also be considered  together with compact localized states to understand the fundamental properties of a flat band completely.
Probing one of those states is equivalent to the observation of the Bloch wave function's singularity of the flat band.
The first experimental attempt to observe non-contractible loop states was done in the photonic Lieb lattice~\cite{xia2018unconventional}.
They considered a finite Lieb lattice with an open boundary condition, and showed that an extended state, called the \textit{line state} is an eigenmode.
Although the line state looks similar to the non-contractible loop state, it is an indirect demonstration of the non-contractible loop state because the stability of the line state relies on the shape of the boundary of the finite system.
Moreover, in the case of the kagome lattice, one cannot find a stable line state for such simple open boundary condition, and we need a more delicate decoration of the boundaries for the line state to become an eigenstate~\cite{ma2020direct}.
In principle, it is impossible to observe the non-contractible loop state directly in a finite system because the non-contractible loop state is well-defined and independent of compact localized states only under the periodic boundary condition.

The periodic boundary conditions in two dimensions are usually hard to be implemented in real space simultaneously.
However, we need just one periodic boundary condition along one direction to stabilize one of the non-contractible loop states of a flat band.
This was realized experimentally by arranging the waveguides of the photonic crystal in an annular disk, called the Corbino disk, where the spatial periodicity is preserved along the azimuthal direction.
The kagome lattice in the Corbino geometry illustrated in Fig.~\ref{fig:observation}(a) was realized in the photonic lattice as shown in Fig.~\ref{fig:observation}(b). 
Then, as in Fig.~\ref{fig:observation}(c) and (d), only the light beam under out-of-phase condition, corresponding to the non-contractible loop state, propagates without dissipation along the transverse direction, while the light beam under in-phase condition is destroyed after the propagation.
The demonstration of the stability of the light beam winding the whole system with out-of-phase configuration is the first direct observation of the real-space topology represented by the non-contractible loop state.

Another topological object manifesting the singularity of the flat band is the robust boundary mode.
The robust boundary mode only requires the open boundary condition, which is more realistic than the periodic boundary condition.
As shown in Fig.~\ref{fig:observation}(e), a finite kagome lattice is synthesized in a photonic lattice.
Then, a state spanning the boundary of the system with out-of-phase configuration is found to be an eigenmode corresponding to the robust boundary mode as presented in Fig.~\ref{fig:observation}(f).
Moreover, the robustness of the robust boundary mode is also checked by showing that the robust boundary mode cannot be cut, although it is slightly deformed, when a defect is placed on the boundary of the system.

\section{Strength of the singularity}\label{strength_of_singularity}

\subsection{Maximum quantum distance}
In topologically non-trivial bands, the bulk topological invariant is nonzero due to a singular behavior of the corresponding Bloch wave function in momentum space.
For example, a Chern band yields a nonzero Chern number as a bulk topological number because the Bloch wave function of this band cannot be determined uniquely and smoothly over the whole Brillouin zone~\cite{kohmoto1985topological}.
In the case of the Weyl semimetal, the nonzero quantized monopole charge of the Weyl point is from the discontinuity of the Bloch wave function at this point~\cite{yang2011quantum}.
In two dimensional band-crossings like the Dirac point of graphene, the Berry phase given by
\ba
\gamma = i\oint_C\langle v_{\b k}| \partial_{\b k} | v_{\b k} \rangle,\label{eq:berry_phase}
\ea
is usually used to determine whether the band-crossing point is topological or not, where $C$ is a closed path enclosing the band-crossing point and $|v_{\b k}\rangle$ is the cell-periodic part of the Bloch wave function~\cite{mikitik1999manifestation,luk2004phase}.
Here, the singularity of the Bloch wave function at the touching point also plays a key role when gamma is nonzero modulo $2\pi$ and quantized.

On the other hand, the band-crossing point of the singular flat band is always topologically trivial due to the flatness condition of the band although the Bloch wave function is singular there~\cite{rhim2020quantum}.
First, the generic singular flat band with a quadratic band crossing does not have a quantized Berry phase (\ref{eq:berry_phase}) because all the Pauli matrices are used in the generic Hamiltonian in (\ref{eq:quad_2by2}).
%
%
Even though the Berry phase is quantized due to some symmetries such as $C_{2z}T$ and $PT$, where $C_{2z}$, $P$, and $T$ are two-fold rotation about the z-axis, inversion, time-reversal symmetries, respectively, the Berry phase of the singular flat band reads $\gamma = 2\pi n$ with integer $n$.
These imply that the band-crossing point of the singular flat band generally does not carry a quantized topological charge, and it is considered topologically trivial.

It was recently proposed that the quantum distance can be used to define a bulk number characterizing the singular band-crossing point of the singular flat band, representing the strength of the singularity~\cite{rhim2020quantum}.
From (\ref{eq:hsqd}), the quantum distance between two Bloch states at momenta $\b k_1$ and $\b k_2$ is given by $d^2_{\b k_1,\b k_2} = 1-|\langle v_{\b k_1} | v_{\b k_2} \rangle |^2$.
Then, the strength of the singularity is defined by the maximum value of the quantum distance between all the possible pairs of Bloch eigenstates of the flat band around the singular point.
Namely,
\ba
d^2_{\mathrm{max}} = \mathrm{max}\left[ d^2_{\b k_1,\b k_2}  \right],
\ea
where $\b k_1$ and $\b k_2$ are momenta close to the band-crossing point.
By definition, the strength of the singularity $d_{\mathrm{max}}$ is valued from 0 to 1.
In the case of the non-singular flat band, the Bloch eigenmode of (\ref{eq:quad_2by2}) is always independent of momentum, so that the quantum distance is zero between an arbitrary pair of momenta.
As a result, $d_\mathrm{max}=0$ for the non-singular flat band.
On the other hand, in the case of the singular flat band, $d_\mathrm{max}$ remains finite even if we consider momenta extremely close to the band-crossing point.
Note that the eigenvector of the Hamiltonian (\ref{eq:quad_2by2}) is independent of $k$, which is the magnitude of the momentum displacement with respect to the band crossing point, because all the dependencies on $k$ can be factored out from this quadratic Hamiltonian.
The general form of the maximum quantum distance of the Hamiltonian (\ref{eq:quad_2by2}) is evaluated as
\ba
d^2_\mathrm{max} = \frac{t_4^2}{-t_2^2 + 4t_1t_3 + 2t_4^2} = 4m_1m_2t_4^2,\label{eq:dmax_formula}
\ea
where $m_1$ and $m_2$ are the minimum and maximum effective masses of the quadratic band touching with the flat band.

The strength of the singularity, $d_\mathrm{max}$, also represents how strong the inter-band coupling between the flat band and parabolic band as explained below by a counting argument~\cite{rhim2020quantum}.
First, the number of independent parameters of the Hamiltonian (\ref{eq:quad_2by2}) is four because five parameters among nine parameters in (\ref{eq:quad_2by2}) are determined by others through the flatness conditions from (\ref{eq:quad_1}) to (\ref{eq:quad_5}).
Since we need three parameters to describe the dispersion relation of the quadratic band of the form $E_\mathrm{quad}(\b k) = a_1 k_x^2 + a_2 k_x k_y + a_3 k_y^2$, and no parameters for the flat band fixed at the zero energy, the remnant one parameter should be related to the inter-band coupling.
Indeed, it was shown that the general flat band model (\ref{eq:quad_2by2}) can be represented by four parameters $M_{xx}$, $M_{xy}$, $M_{yy}$, and $d_\mathrm{max}$, where $M_{\alpha\beta}^{-1}= \partial_{k_\alpha}\partial_{k_\beta} E_\mathrm{quad}(\b k)$ is the mass tensor for the parabolic band.
This implies that $d_\mathrm{max}$ is very natural inter-band coupling parameter of the two dimensional flat band model with a quadratic band touching.

As an example, let us consider the kagome lattice.
Its low energy effective Hamiltonian around the quadratic band-crossing is given by
\ba
\mathcal{H}_\mathrm{kagome}^\mathrm{eff} = \bpm k_x^2 & i k_x k_y \\ -ik_x k_y & k_y^2\epm, \label{eq:kagome_ham_eff}
\ea
where the flat band's energy is shifted to zero~\cite{rhim2019classification}.
The eigenvector of the flat band of this model is evaluated as
\ba
\b v_\mathrm{fb} =\frac{1}{\sqrt{k_x^2+k_y^2}} \bpm -ik_y \\ k_x\epm = \bpm -i\sin\theta \\ \cos\theta \epm,
\ea
where $\theta$ is the polar angle with respect to the band-crossing point at $\b k = 0$.
Then, the quantum distance between two eigenvectors at $\theta_1$ and $\theta_2$ is obtained as
\ba
d^2_{\theta_1,\theta_2} = 1- |\langle \b v_\mathrm{fb}(\theta_1) |\b v_\mathrm{fb}(\theta_2) \rangle |^2 = 1-\cos^2(\theta_1-\theta_2).
\ea
This gives us the maximum quantum distance $d_\mathrm{max} =1$ when $\theta_1-\theta_2=\pi/2$, which is consistent with the formula (\ref{eq:dmax_formula}) because the band parameters for $\mathcal{H}_\mathrm{kagome}^\mathrm{eff}$ are given by $t_1=b_1=-t_3=b_3=1/2$, $t_4=-1$, and zero for other parameters.
Since the quantum distance is ranging from 0 to 1 by definition, the kagome lattice's flat band corresponds to the maximally singular case.

\begin{figure}
	\begin{center}
		\includegraphics[width=1\columnwidth]{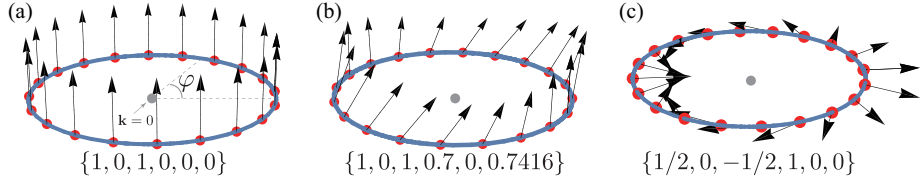}
	\end{center}
	\caption{ Pseudospin structures for various sets of the band parameters $\{t_1,t_2,t_3,t_4,t_5,t_6\}$.
	}
	\label{fig:pseudospin}
\end{figure}

\subsection{Maximum pseudospin canting angle}
The singular band-crossing point of the flat band can be also characterized by the canting structure of the pseudospin given by
\ba
\b s(\b k) = \langle v_{\b k} | \boldsymbol\sigma | v_{\b k} \rangle =  \sum_{\alpha=x,y,z} \hat{\alpha} \frac{f_\alpha(\b k)}{\sqrt{f_x(\b k)^2+f_y(\b k)^2+f_z(\b k)^2}},
\ea
where $| v_{\b k} \rangle$ is the eigenvector of the flat band and $f_\alpha(\b k)$ is the coefficient of $\sigma_\alpha$ in (\ref{eq:quad_2by2}) given by $f_x(\b k) = t_6k_y^2$, $f_y(\b k) =t_4k_xk_y +t_5k_y^2 $, and $f_z(\b k) = t_1 k_x^2 + t_2 k_x k_y + t_3 k_y^2$~\cite{rhim2020quantum}.
One can represent the pseudospin on a surface of a unit sphere because $f_x(\b k)$, $f_y(\b k)$, and $f_z(\b k)$ are all in quadratic order of $k=|\b k|$ so that the size of the pseudospin is constant.
Pseudospin texture is drawn on a circular path encircling the band-crossing point as illustrated in Fig.~\ref{fig:pseudospin}.
Since the eigenvector of (\ref{eq:quad_2by2}) is independent of $k$, the pseudospin structure is also independent of the size of circle.
In the case of the non-singular flat band, the pseudospins are aligned to each other as shown in Fig.~\ref{fig:pseudospin}(a).
As a result, the pseudospin is well-defined even at the band-crossing point.
However, in the case of the singular flat band, we have a canting structure of the pseudospins around the band-crossing point as illustrated in Fig.~\ref{fig:pseudospin}(b) and (c).
This implies that the pseudospin cannot be fixed uniquely at the touching point, which is consistent with the fact that the corresponding Bloch eigenstate is not uniquely defined there.

It was shown that the quantum distance between two Bloch states at $\b k_1$ and $\b k_2$ can be represented by the relative angle $\Delta\theta$ between two pseudospins at the same momenta as follows.
\ba
\Delta\theta(\b k_1,\b k_2) = 2\sin^{-1} d_{\b k_1,\b k_2},
\ea
which represents the one-to-one correspondence between those two quantities.
Therefore, one can use the maximum canting angle $\Delta\theta_\mathrm{max} = \mathrm{max}\Delta\theta(\b k_1,\b k_2)$ as the strength of the singularity of the singular flat band instead of the maximum quantum distance $d_\mathrm{max}$.
According to the maximum canting angle, the value of the strength of the singularity ranges from 0 (non-singular flat band) to $\pi$ (maximally singular flat band).
In general, due to the topological triviality, the pseudospin texture of the singular flat band does not exhibit any winding structure except the maximally singular case ($\Delta\theta_\mathrm{max} = \pi$ or $d_\mathrm{max}=1$), where we have a winding number 2 of the pseudospin around the band-crossing point.
However, the singular flat band can be characterized by the pseudospin canting structure or the finite maximum quantum distance, which can be manifested in reality via Landau level structure as explained in the next section.

Let us consider again the kagome lattice as an example.
For convenience, we use a unitary transformed Hamiltonian $\mathcal{\tilde{H}}_\mathrm{kagome}^\mathrm{eff}  = \mathcal{U}^\dag\mathcal{H}_\mathrm{kagome}^\mathrm{eff} \mathcal{U}$, where $\mathcal{U} = (\sigma_0 - i\sigma_y)/\sqrt{2}$.
Then, $\mathcal{\tilde{H}}_\mathrm{kagome}^\mathrm{eff} $ is described by $f_x(\b k) = (-k_x^2+k_y^2)/2$, $f_y(\b k) = -k_xk_y$, $f_z(\b k) =0$, and $f_0(\b k) = (k_x^2+k_y^2)/2$.
As a result, the pseudospin formula is given by
\ba
\b s(\theta) = -\cos2\theta~\hat{x} - \sin2\theta~\hat{y},
\ea
which shows double winding around the band-crossing point, and $\Delta_\mathrm{max}=\pi$.

\section{Anomalous Landau levels of singular flat bands}\label{sec:anomalous_landau_level}

\subsection{Onsager's semiclassical theory}
In solids, the band structure of electrons under a magnetic field, namely the Landau level structure, can be understood intuitively as well as quantitatively by using the Onsager's semiclassical quantization rule described by
\ba
S_0(\varepsilon) =\frac{2\pi e B}{\hbar}\left(n +\frac{1}{2}-\frac{\gamma_{\varepsilon,B}}{2\pi}\right), 
\label{eq:Onsager}
\ea
where $\varepsilon$ represents energy, $S_0(\varepsilon)$ is the area of a closed path at the energy $\varepsilon$ on a band, $B$ is the magnetic field, $e$ is the electric charge, $h=2\pi\hbar$ is the Planck constant, $n$ is an integer corresponding to the Landau level index, and $\gamma_{\varepsilon,B}$ is the quantum correction due to Berry phase, magnetic susceptibility, and higher order magnetic responses~\cite{onsager1952interpretation,roth1966semiclassical,mikitik1999manifestation,gao2017zero,fuchs2018landau}.
Only the energy values of $\varepsilon$ satisfying the above equation are allowed as eigenenergies, and they are quantized because the Landau level index $n$ is discretized.
The key idea of this scheme is that one can predict the Landau level structure from the information obtained before applying magnetic field such as the band dispersion and geometric properties of the Bloch wave function.
Conversely, one can extract the geometric properties of the Bloch wave function, such as the Berry phase, from the Landau level structure via the above formula.
Let us consider graphene as an example.
Denoting the Fermi velocity around its Dirac point by $v_F$, we have $S_0(\varepsilon) = \pi (\varepsilon/\hbar v_F)^2$ and $\gamma_{\varepsilon,B} = \pi$.
This leads to the well-known Landau level quantization $\varepsilon = v_F (2e\hbar B n)^{1/2}$.

\begin{figure}
	\begin{center}
		\includegraphics[width=1\columnwidth]{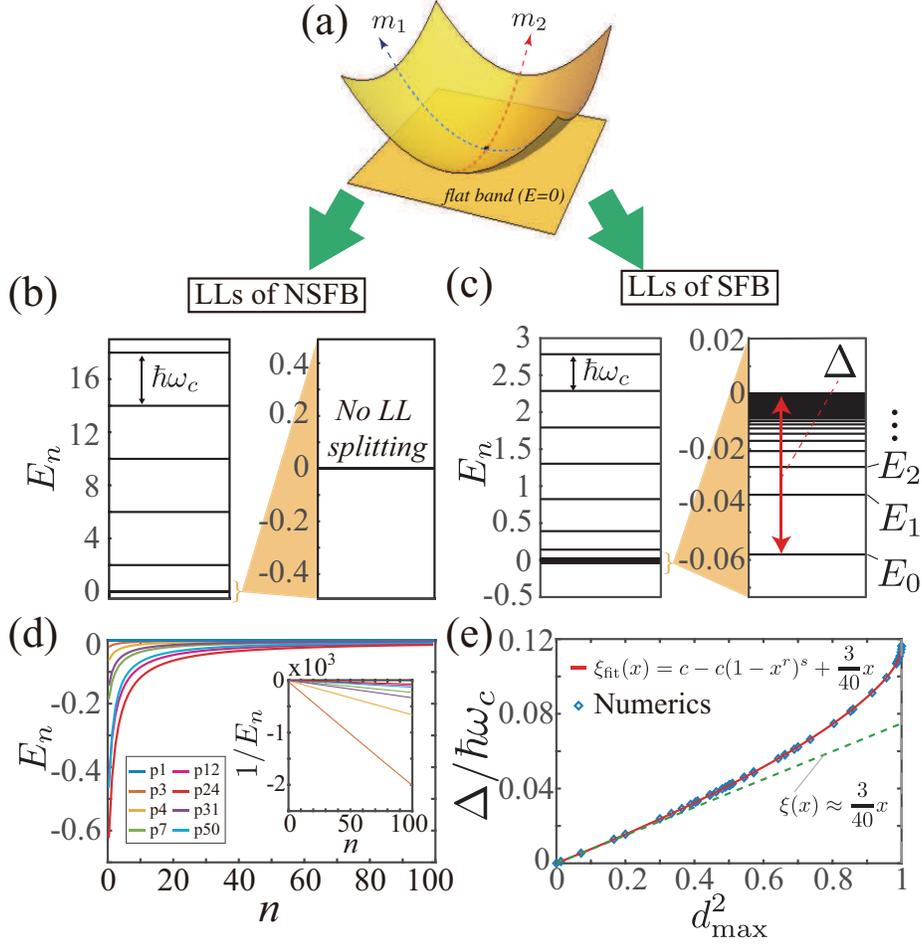}
	\end{center}
	\caption{ (a) A flat band and a parabolic band crossing each other at a point. $m_1$ and $m_2$ represent the minimum and maximum effective masses. Generic shape of the Landau level structure of (b) the non-singular flat band and (c) the singular flat band. In (c), $\Delta$ is the Landau level spreading of the flat band. (d) $1/n$ dependence of the anomalous Landau levels corresponding to the singular flat band. (e) The ratio $\Delta/\hbar\omega_c$ is plotted as a function of $d^2_\mathrm{max}$. Diamond symbols are the numerical results from the tight binding lattice model, and the red solid curve is the fitting function for the data.
	}
	\label{fig:landau_level}
\end{figure}

\subsection{Landau level spreading of the singular flat band with a quadratic band-crossing}\label{subsec:landau_level_spreading}
The Onsager's scheme cannot be applied to the flat bands because $S_0(\varepsilon)$ is not well-defined in this case.
Namely, one can have infinitely many choices of the closed path at the energy of the flat band, and therefore $S_0(\varepsilon)$ is not single-valued.
What we can naively infer from the fact that the effective mass $m^*$ is infinite in the flat band is that the cyclotron energy ($\hbar\omega_c = \hbar e B/m^*$) is vanishing and all the Landau levels are trivially developed at the same energy of the flat band.
However, it was recently shown that the Landau levels of a flat band are nontrivial and determined by a completely different scheme involving the quantum distance of the Bloch states of the flat band around the singular momentum, where the flat band is touching with another quadratically, as described below~\cite{rhim2020quantum}.

To consider the effect of the magnetic field on the flat band, the transformations $k_x\rightarrow (a+a^\dag)/\sqrt{2}l_B$ and $k_y\rightarrow i(a-a^\dag)/\sqrt{2}l_B$ are applied to the quadratic band-crossing flat band Hamiltonian in (\ref{eq:quad_2by2}), where $a$ and $a^\dag$ are the annihilation and creation operators, respectively satisfying $[a,a^\dag]=1$.
To ensure the Hermiticity, $k_xk_y$ is symmetrized by replacing it with $(k_xk_y + k_yk_x)/2$.
Then, each element of (\ref{eq:quad_2by2}) becomes an operator.
This Hamiltonian is solved by applying a trial wave function of the form
\begin{align}
|\psi \rangle = \sum_{n=0}^\infty \bpm C_n \\ D_n \epm | u_n\rangle,\label{eq:state}
\end{align}
where $| u_n\rangle$ is the normalized eigenstate of the harmonic oscillator satisfying $a| u_n\rangle = \sqrt{n}| u_{n-1}\rangle$ and $a^\dag | u_n\rangle = \sqrt{n+1}| u_{n+1}\rangle$, and $C_n$ and $D_n$ are complex coefficients.

The generic structure of the Landau levels of the flat band model (\ref{eq:quad_2by2}) obtained by the procedure described above is shown in Fig.~\ref{fig:landau_level}.
The most intriguing feature is that the Landau levels are developed at the energies, where the density of states was zero before applying the magnetic field.
This behavior is completely anomalous from the semiclassical point of view because the semiclassical quantization equation (\ref{eq:Onsager}) does not have a solution when the energy $\varepsilon$ is in the band gap region.
This phenomenon is called the \textit{Landau level spreading} of a flat band~\cite{rhim2020quantum}.
In the case shown in Fig.~\ref{fig:landau_level}(a), the energy of the flat band is lower than that of the parabolic band, the energy difference between the flat band and the lowest Landau level defines the magnitude of the Landau level spreading.
On the other hand, if the flat band is above the hole-like quadratic band, the Landau level spreading is the energy of the highest Landau level with respect to the flat band's energy.

While the Landau level spreading appears only in the singular flat bands, its magnitude can be shown to be determined by the maximum quantum distance $d_\mathrm{max}$ of the Bloch eigenstate before applying magnetic field.
More specifically, as shown in Fig.~\ref{fig:landau_level}(e), the ratio $\Delta/\hbar\omega_c$ is a simple monotonic function of $d_\mathrm{max}$ given by
\ba
\frac{\Delta}{\hbar\omega} = \xi(d^2_\mathrm{max}), \label{eq:LL_spreading}
\ea
where $\Delta$ is the Landau level spreading and $\omega_c = eB/\sqrt{m_1m_2}$ is the cyclotron frequency of the parabolic band.
It is important to note that $\Delta/\hbar\omega_c$ is only a function of $d_\mathrm{max}$ regardless of an arbitrary choice of band parameters of the model Hamiltonian in (\ref{eq:quad_2by2}).
This implies that the quantum distance plays the key role in the Landau level structure of the flat band, and the strength of the singularity of the band-crossing point manifests itself as the Landau level spreading.
In fact, it was rigorously shown that two Hamiltonians with distinct sets of band parameters are equivalent to each other within a scale factor if they share the same $d_\mathrm{max}$.
The scale factor is the ratio between the cyclotron energies of the parabolic bands of those two Hamiltonians.
Therefore, all the singular flat band systems with the same $d_\mathrm{max}$ share the same value of $\Delta/\hbar\omega_c$.
Moreover, the Landau level spreading can be understood as a result of the level repulsion between Landau levels originating from the flat band and the parabolic band.
Indeed, the coupling strength between Landau levels from the flat and the parabolic bands, which is relevant to the level repulsion, was shown to be proportional to $d_\mathrm{max}$ in the weakly singular limit ($d_\mathrm{max} \ll 1$).

While the explicit form of $\xi(x)$ is unavailable, its approximate form, which fits the numerical data accurately, was found to be
\ba
\xi_\mathrm{fit}(x) = c-c(1-x^r)^s +\frac{3}{40}x,
\ea
where $c=0.041$, $r=2.344$, and $s=0.487$.
This was obtained from several analytic results such as $\xi(d_\mathrm{max}^2 \rightarrow 0) = (3/40)d_\mathrm{max}^2$ and $\xi(d_\mathrm{max}^2 \rightarrow 1) = (2\sqrt{3}-3)/4$.
By using this formula, one can extract the geometric information, namely the maximum quantum distance $d_\mathrm{max}$, of the singular flat band around the band-crossing point from the Landau level structure.

Another intriguing feature of the Landau level structure of the singular flat band with a quadratic band-crossing is that the Landau levels are inversely proportional to the Landau level index $n$ close to the flat band's energy as shown in Fig.~\ref{fig:landau_level}(d). 
Here, the energy of the flat band is assume to be zero without loss of generality.
As an example, in the case of the maximally singular ($d_\mathrm{max}=1$) model in (\ref{eq:kagome_ham_eff}), the explicit formula for the Landau levels corresponding to the singular flat band is given by
\ba
E^\mathrm{LL}_\mathrm{fb}(n) = \frac{1}{2l_\mathrm{B}^2}\left( 2n+3 - 2\sqrt{(n+1)(n+2)+1} \right),
\ea
where $l_\mathrm{B} = \sqrt{\hbar/eB}$ is the magnetic length.
Then, close to the flat band's energy ($E=0$), namely for $n\gg 1$, the above formula is approximated to
\ba
E^\mathrm{LL}_\mathrm{fb}(n) \approx - \frac{3}{8l^2_\mathrm{B}}\frac{1}{n}.
\ea
The same behavior appears in the weakly singular cases ($d_\mathrm{max}\ll1$) too.
One of the simplest example of the weakly singular flat band models is obtained by the band parameters $t_1=b_1=t_3$, $t_2=b_2=t_5=0$, $t_6=t_4\sqrt{4t_1^2+t_4^2}/(2t_1)$, and $b_3=\sqrt{t_1^2 + t_4^2(4t_1^2+t_4^2)/(4t_1^2)}$, where $t_1$ and $t_4$ are free parameters with $t_4 \ll t_1$.
In this case, we have $d_\mathrm{max} \approx t_4/(2t_1)$.
According to the level repulsion mechanism mentioned in the previous paragraph, it was shown that the approximate formula for the Landau levels corresponding to the flat band is given by
\ba
E^\mathrm{LL}_\mathrm{fb}(n) \approx -\frac{3t_4^2}{16 t_1^2 l^2_\mathrm{B}} \frac{1}{n}.
\ea

\begin{figure}
	\begin{center}
		\includegraphics[width=1\columnwidth]{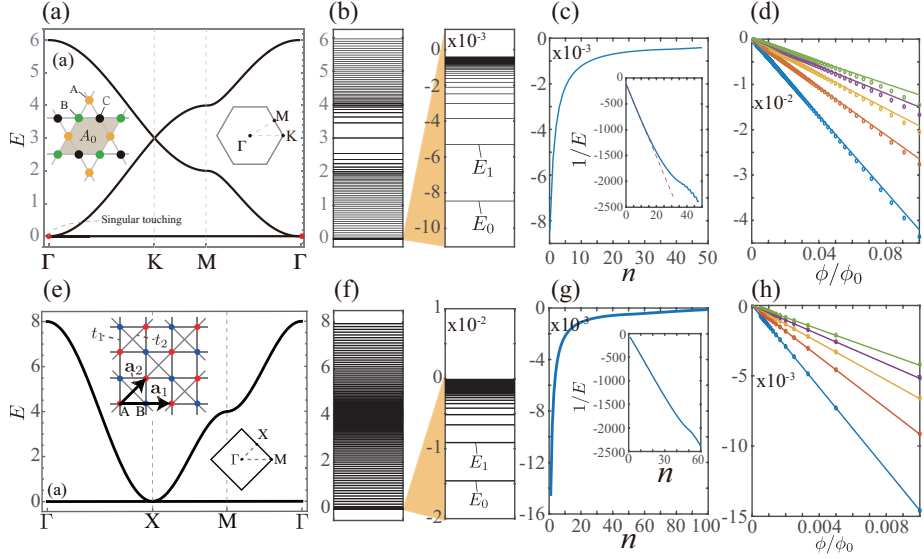}
	\end{center}
	\caption{ Lattice structures and band structures of the kagome and Lieb lattices are shown in (a) and (e) respectively. Landau level structures of those models are plotted in (b) and (f). On the right-hand side panels, the Landau levels around the flat band's energy are highlighted. In (c) and (g), the $1/n$ dependency of the Landau level energies around the zero energy are exhibited. It is shown that the Landau levels obtained from the tight binding lattice model (circular symbols) are consistent with the continuum results (solid curves) from (\ref{eq:LL_spreading}) in (d) and (h).
	}
	\label{fig:lattice_landau}
\end{figure}

\subsection{Lattice models}
The results in Sec.~\ref{subsec:landau_level_spreading} obtained from the continuum Hamiltonian (\ref{eq:quad_2by2}) were successfully applied to several lattice models such as the kagome and the checkerboard lattice models~\cite{rhim2020quantum}.
Such lattice models are described by the tight binding Hamiltonians of the form
\ba
\mathcal{H}_\mathrm{lattice} = \sum_{i,j} t_{ij} c^\dag_i c_j,
\ea
where $t_{ij}$ is the hopping parameter between the $i$-th and $j$-th sites, and $c_i$ represents the annihilation operator at the $i$-th site. 
The effect of the magnetic field on this system can be considered by the Peierls substitution given by
\ba
t_{ij} \rightarrow t_{ij} e^{i2\pi\frac{1}{\phi_0} \int_i^j \b A\cdot d\b l },
\ea
where $\b A$ is the vector potential corresponding to the applied magnetic field, and $\phi_0 = h/e$ is the magnetic flux quantum~\cite{hofstadter1976energy}.
Denoting the magnetic flux penetrating the unit cell before applying the magnetic field by $\phi$, we only consider the cases, where $\phi/\phi_0 = p/q$ with natural numbers $p$ and $q$.

The Landau levels of the kagome lattice with only the nearest neighbor hopping processes are shown in Fig.~\ref{fig:lattice_landau}(b), and those of the checkerboard lattice model with the nearest and the next nearest hopping processes are plotted in Fig.~\ref{fig:lattice_landau}(f).
Most importantly, the Landau level spreading phenomenon of the flat band discussed in the continuum model analysis appears in the Landau level structures of both lattice models too.
According to the continuum result in (\ref{eq:LL_spreading}), the Landau level spreading should be $\Delta = \hbar\omega_c (2\sqrt{3}-3)/4$, where $\hbar\omega_c$ is the Landau level spacing of the parabolic band, because the singular flat bands of the kagome and checkerboard lattices yield $d_\mathrm{max}=1$ around their band-crossing points.
As shown in Fig.~\ref{fig:lattice_landau}(d) and (h), this continuum results (solid lines) are consistent with the numerical data obtained from the lattice models (circular symbols).
Moreover, the $1/n$ behavior of the Landau levels of the continuum model is clearly shown in the Landau levels of the lattice models too as plotted in Fig.~\ref{fig:lattice_landau}(c) and (g).
However, this trend breaks down for large $n$'s due to the lattice effect.
A distinguishing feature of the results of the lattice model is that there are finite number of Landau levels while there are infinite ones in the continuum model.
In the lattice model, after the Peierls substitution, we have an enlarged unit cell, called the magnetic unit cell.
Let us assume that the magnetic unit cell is $Q$ times larger than the original unit cell, where $Q$ is a natural number determined by the flux number $\phi/\phi_0$.
Then, to preserve the total number of states before and after the application of the magnetic field, we have $Q$ magnetic bands, corresponding to the Landau levels, per each original energy band.
In those two lattice models under the magnetic field described by $\phi/\phi_0 = 1/q$, one can find a gauge choice where the magnetic unit cell is $q$ times larger than the original unit cell, and we have $q$ magnetic bands corresponding to their flat band.
It was shown that only $q-1$ magnetic bands are developed in the band gap region below the flat band, and one remnant magnetic band is above it.

\subsection{Realistic systems}
In real materials, one cannot have a perfectly flat band because the hopping parameters of such systems cannot be fine-tuned as in the toy models treated in the previous subsections.
For instance, while a perfectly flat band appears in the kagome lattice when we consider only the nearest neighbor hopping processes, it becomes dispersive as soon as we include long range hopping processes or the spin-orbit coupling (SOC) which exist in the realistic situations.
Let us denote the first and second nearest neighbor hopping parameters by $t_1$ and $t_2$, and the SOC strength by $\lambda$ respectively.
Then, we have a nearly flat band when the later two processes are perturbatively small, namely $t_2,~\lambda \ll t_1$.
There are several candidate materials with kagome-like structures hosting nearly flat bands belonging to different regimes in the parameter space of $t_2/t_1$ and $\lambda/t_1$ as follows.
First, the carbon networks such as the cyclic graphene, cyclic graphyne, and cyclic graphdiyne, belong to the case where $\lambda \approx 0$ because the SOC can be neglected due to the low atomic number of carbon~\cite{rhim2020quantum,chen2018ferromagnetism,you2019flat}.
In this case, the gap opening at the band-corssing point is almost negligible while the flat band obtains a small dispersion due to the finite value of $t_2/t_1$.
Second, the nearly commensurate-charge density wave (NC-CDW) phase of 1T-TaS$_2$ corresponds to the regime $t_2/t_1 \approx 0$ because the second nearest neighbor hopping processes as well as the longer range hopping parameters are exponentially suppressed due to the large insulating domains~\cite{lee2020stable}.
Third, in the trans-Au-THTAP, both $t_2/t_1$ and $\lambda$ are non-negligible although they are small enough to have a nearly flat band~\cite{yamada2016first}.
In the last two cases, the singularity of the unperturbed flat band is completely removed by the energy gap induced by the SOC.
However, this singularity is still manifested in the Landau level structure of the nearly flat band as shown in the following formula for the zeroth Landau level of the nearly flat band:
\ba
E^{\pm}_0 \approx 2(t_1+t_2)+a_0\frac{\phi}{\phi_0} \pm \sqrt{a_1 \frac{\phi^2}{\phi_0^2} - \frac{8\lambda a_0}{\sqrt{3}}\frac{\phi}{\phi_0} + 12\lambda^2},\label{eq:tat_LL_0}
\ea
where $a_0 = -\sqrt{3}\pi (t_1/2 +2t_2)$, and $a_1 = \pi^2 (t_1^2 +8 t_2^2)$~\cite{rhim2020quantum}.
Here, the zeroth Landau level $E^{+}_0$ ($E^{-}_0$) corresponds to the lowest (highest) Landau level of the flat band touching with an electron-like (hole-like) parabolic band.
The electron-like (hole-like) parabolic band can be obtained by the negative (positive) value of $t_1$.
This formula can be applied to all the three cases discussed in the above, and it was shown that the maximum quantum distance can be extracted successfully from this in the strong magnetic field limit.
More specifically, the Landau level spreading of the unperturbed flat band competes with the energy scale of the band width of the nearly flat band.
As a result, under very weak magnetic field, the band width of the nearly flat band dominates the landau level spreading so that the Onsager's semiclassical scheme can be successfully applied to obtain the Landau levels.
However, if the magnetic field is strong enough so that the Landau level spreading is much larger than the band width, the Landau level spreading of the nearly flat band can be clearly observed.

\subsection{Diverging orbital magnetic susceptibility}
At zero temperature, the orbital magnetic susceptibility is defined by
\ba
\chi_\mathrm{orb} = -\lim_{B\rightarrow 0} \frac{d^2 E_\mathrm{tot}}{d B^2},
\ea
where $E_\mathrm{tot}$ is the total energy per unit area of the system under the magnetic field.
As a representative example, the diamagnetic divergence of the orbital magnetic susceptibility of graphene can be understood from this definition as follows.
The Landau levels of graphene are proportional to $\sqrt{B}$.
Since the Landau level degeneracy is proportional to $B$, the total energy below the Dirac point shows $B\sqrt{B}$ dependence, which results in $1/\sqrt{B}$ divergence in the weak magnetic field limit.
Therefore, the peculiar divergence of the orbital magnetic susceptibility of graphene is due to the $\sqrt{B}$ dependence of the Landau levels~\cite{ominato2013orbital}.

On the other hand, the Landau levels of the singular flat band are proportional to $B$ as in the conventional cases.
Instead, they show the intriguing $1/n$ dependence~\cite{rhim2020quantum}.
As a result, the total energy for a fully occupied singular flat band is given by
\ba
E_\mathrm{tot} \propto \frac{1}{2\pi l_\mathrm{B}^2} \sum_{n=1}^q \frac{1}{n  l_\mathrm{B}^2} \propto B^2\ln B,
\ea
where we assume that the parabolic band touching with the singular flat band at the zero energy is electron-like, and the magnetic flux per unit cell is $\phi/\phi_0 = 1/q$ with $q\gg 1$.
This form of the total energy leads to the logarithmically divergent paramagnetism, namely $\chi_\mathrm{orb} \propto -\ln B + \mathrm{const}$.
On the contrary, if the parabolic band is hole-like, the singular flat band gives the diverging diamagnetism.
This logarithmic divergence of the orbital magnetic susceptibility can be seen even for the partially filled flat band.

The fate of the orbital magnetic susceptibility when the flat band gains a small dispersion due to a finite SOC is also important~\cite{rhim2020quantum}.
In this case, the singular flat band becomes an insulator due to the gap opening at the band-crossing point.
When the resulting nearly flat band is fully occupied, namely the Fermi level is in the gap between the flat band and the parabolic band, it was found that the orbital magnetic susceptibility is no more divergent.
However, the orbital magnetic susceptibility still remains finite even though there is no state in the gap, and its value could be large for the small enough SOC.

\subsection{Landau level broadening due to disorder}
When the system is disordered, its Landau levels cannot be described by the Dirac delta function type density of states anymore, and are broadened.
It was shown that the Landau level broadening appears differently in the non-singular flat band and singular flat band, and the quantum distance plays a central role~\cite{rhim2020quantum}.
Quantitatively, the Landau level broadening $\Gamma$ is defined as the energy width of the density of states of the broadened Landau level at its half-maximum.
In the recent work~\cite{rhim2020quantum}, an onsite disorder potential given by $\mathcal{V}_\mathrm{disorder} = \rho \sum_i \epsilon_i c^\dag_i c_i$ is considered, where $i$ is the site index, $\epsilon_i$ is a random number ranging from $-0.5$ to $0.5$, and $\rho$ is the disorder strength.
First, in the case of the non-singular flat band, where all the Landau levels corresponding to the flat band are degenerate at the energy of the flat band, the Landau level broadening $\Gamma$ is proportional to $\rho$, and almost independent of the magnetic field.
This is a different behavior from the usual Landau level broadening of parabolic dispersions within the Born approximation given by $\Gamma =(\hbar^2\omega_c/2\pi\tau_0)^{1/2}\propto \sqrt{B}$, where $\tau_0$ is the relaxation time in the absence of the magnetic field.
On the other hand, in the case of the singular flat band, all the Landau levels spread into the energy gap are non-degenerate, and the Landau level broadening of each of them should be investigated.
In the weakly singular cases, it was found that the Landau level broadening of the singular flat band is due to the inter-band coupling between the singular flat band and the parabolic band touching with it.
Indeed, the approximate formula for the Landau level broadening is given by
\ba
\Gamma_\mathrm{fb} = \frac{d_\mathrm{max}^2}{4}\sqrt{\frac{\hbar^2\omega_c}{2\pi\tau_0}},\label{eq:landau_level_broadening}
\ea
where $\hbar\omega_c$ is the cyclotron energy of the quadratic band.
Note that $d_\mathrm{max}$ represents the strength of the inter-band coupling as discussed in Sec.~\ref{subsec:landau_level_spreading}.
While the $\sqrt{B}$ behavior of the Born approximation is recovered, it is not the intrinsic property of the singular flat band.
Instead, it arises from the Landau level broadening  ($\sim \sqrt{B}$) of the parabolic band, which appears in Eq.(\ref{eq:landau_level_broadening}) through the inter-band coupling due to the singularity at the band crossing point.
To observe the Landau level spreading in experiments, it should be much larger than the Landau level broadening and it was shown in \cite{rhim2020quantum} that the critical magnetic field and the disorder strength are within the experimentally accessible range.
%

\section{Conclusions and outlook}\label{sec:outlook}

We have reviewed the theoretical and experimental progress of the fundamental understanding of singular flat band systems.
A singular flat band has a quite simple band structure, composed of a flat band crossing with another dispersive band at a momentum. However, the singularity of the flat band's Bloch states gives unexpected physical outcomes. The presence of partially extended eigenstates, such as the non-contractible loop states and the robust boundary modes. More surprisingly, the singularity is deeply rooted from the quantum distance of Bloch wave functions around the band crossing point. Such intriguing geometric properties of the singular flat band eventually leads to the unusual Landau level spreading, which goes beyond the paradigm of Onsager's semi-classical theory.

We believe that singular flat bands are an ideal new platform where intriguing geometric properties of Bloch states associated with quantum metric or quantum distance can be investigated. Although it is trivial in view of the conventional band topology, it provides an avenue to explore the geometry of quantum states from a different perspective, beyond the physics of Berry curvature.
In this respect, recent experimental discovery of various materials hosting nearly flat bands, such as CoSn~\cite{kang2020topological,liu2020orbital} and FeSn~\cite{kang2020dirac}, is quite encouraging.
We note that while the most actively studied flat band system recently is the twisted bilayer graphene, the nearly flat bands of the twisted bilayer graphene just belong to the non-singular flat band according to the classification scheme in this review~\cite{rhim2020quantum}.
On the other hand, the flat band of CoSn and FeSn, arising from the underlying kagome lattice structure, belongs to the category of singular flat bands.
We propose that the flat band materials based on kagome lattice structure would provide new opportunities to reveal the geometric properties related to the nonzero quantum distance predicted in this paper.
It is worth noting that, unlike the physics of Berry curvature and Berry phase, which is already well-established, relatively less attention has been paid to the physical significance of the quantum metric and quantum distance. 
We believe that the study of singular flat bands would pave the way for the complete understanding of geometric properties of Bloch wave functions in condensed matter physics, and it could lay groundwork to figure out the underlying mechanism of the strongly correlated physics in flat band system, such as the superconducivity and magnetism.

\section{Disclosure statement}

No potential conflict of interest was reported by the authors.

\section{Funding}

J.W.R. was supported by the Institute for Basic Science in Korea (Grant No. IBS-R009-D1). 
B.J.Y. was supported by the Institute for Basic Science in Korea (Grant No. IBS-R009-D1) and Basic Science Research Program through
the National Research Foundation of Korea (NRF) (Grant
No. 0426-20200003), and the US Army Research Office under Grant
No. W911NF-18-1-0137.

\section{References}

\bibliographystyle{unsrt}
\bibliography{Reference.bib}

\end{document}